\documentclass{aa}
\usepackage{txfonts}
\usepackage{booktabs}
\usepackage{graphicx,natbib}
\bibpunct{(}{)}{;}{a}{}{,} 
\newcommand{\ltsima} {$\; \buildrel < \over \sim \;$}  
\newcommand{\gtsima} {$\; \buildrel > \over \sim \;$}  
\newcommand{\lta} {\lower.5ex\hbox{\ltsima}}  
\newcommand{\gta} {\lower.5ex\hbox{\gtsima}}  
\newcommand{\Ha} {H$\alpha$}

\usepackage{color}

\begin{document}

\title{The large-scale environment of 3CR radio galaxies at z<0.3}

\author{Samantha Casadei\inst{1,2}, Alessandro Capetti\inst{2}, Claudia M. Raiteri\inst{2}, Francesco Massaro\inst{1}}
\institute{
  Dipartimento di Fisica, Universit\`a degli Studi di
  Torino, Via Pietro Giuria 1, 10125 (Torino), Italy
   \and
  INAF - Osservatorio
  Astrofisico di Torino, Via Osservatorio 20, I-10025 Pino Torinese,
  Italy}

  \abstract{The question of whether and how the properties of
    radio galaxies (RGs) are connected with the large-scale environment is
    still an open issue.  For this work we measured the large-scale
    galaxies' density around RGs present in the revised Third Cambridge Catalog of radio sources (3CR) with $0.02 < z < 0.3$.  The goal is to determine whether the accretion mode and morphology of RGs are related to the richness of the environment.

  We considered RGs at $0.05 < z < 0.3$ for a comparison between
  optical spectroscopic classes, and those within $0.02 < z < 0.1$ to
  study the differences between the radio morphological types.
  Photometric data from the Panoramic Survey Telescope \& Rapid
  Response System (Pan-STARRS) survey were used to search for "red
  sequences" within an area of 500 kpc of radius around each RG.
  
We find that 1) RGs span over a large range of local galaxies' density, from isolated sources to those in rich environments, 2) the richness distributions of the various classes are not statistically different, and 3) the radio luminosity is not connected with the source environment. Our results suggest that the RG properties are independent of the local galaxies density, which is in agreement with some previous analyses, but contrasting with other studies. We discuss the possible origin of this discrepancy. An analysis of a larger sample is needed to put out results on a stronger statistical basis.}
  
  \titlerunning{The environment of 3C radio-galaxies with $z$=0.3} \authorrunning{Casadei et al.}
\maketitle
  
  \section{Introduction}
\label{intro}

Studies on active galactic nuclei (AGN) have a major role in astrophysics. Their energetic processes are believed to be fundamental in the evolution of their host galaxy (e.g., \citealt{ferrarese00}, \citealt{haring04}, \citealt{vanden06}) and the
general environment they inhabit (e.g., \citealt{silverman09}, \citealt{kollatschny12}).
In this context, radio galaxies (RGs) are an ideal laboratory to investigate the link between the activity of the central engine and the large-scale structure in which they live. In particular, it is important to establish whether the properties of RGs, from their radio morphology to their accretion efficiency, are linked to their environment.

Extended extragalactic radio sources can be classified as \citet{fanaroff74}  edge-darkened FR~I, when they are dominated by emission of two-sided jets, or edge-brightened FR~II if they are dominated by two lobes. Although it has been proposed that this dichotomy arises from differences in the properties of the central engine (e.g., \citealt{baum95}, \citealt{zirbel95}), it might also be connected with their environment (e.g., \citealt{gopal01}). 

In addition to the radio dichotomy, optical studies have discovered different AGN types
considering their spectroscopic appearance. 
\citet{laing94}, following the original suggestion by \citet{hine79}, found that FR~IIs can be
put into two subclasses. They proposed a separation into
high excitation galaxies (HEGs) -- which are defined as galaxies with
[O~III]$/$\Ha$>0.2$ and an equivalent width (EW) of [O~III] $>$ 3 \AA\ --
and low excitation galaxies (LEGs). \citet{tadhunter98} found a
similar result from an optical spectroscopic study of the 2Jy sample, in which a subclass of weak-line RGs (sources with an EW of [O~III])$<10$\AA) stands out due to a low ratio between emission line
and radio luminosities as well as in the [O~III]/[O~II] line ratio.
These early results have been confirmed by \citet{buttiglione10,buttiglione11} from the study of 3C sources with $z<0.3$ and by \citet{capetti22} for sources up to z$\sim$ 0.8.
The difference between LEGs and HEGs is ascribed to the intrinsic
efficiency of the accretion onto the super massive black hole (SMBH): HEGs are characterized by a high Eddington
ratio (i.e., $L_{acc}/L_{Edd} > 0.01$) and LEGs by inefficient
accretion \citep{buttiglione10}.  
Along this main classification, a further class called broad line objects (BLOs) includes the sources in which the optical spectrum shows broad permitted lines. 
However, an AGN classification does not always arise from intrinsic differences. In fact, anisotropy effects (e.g., \citealt{antonucci93}, \citealt{urry95}) can be fundamental in changing the spectral characteristics of an AGN. Hence, their diversities are often just due to an orientation with respect to our line of sight, rather than a physical phenomenon. For example,
the latter class mentioned (i.e., BLOs) has accretion properties and spectral narrow emission lines typical of HEGs where the lack of broad emission lines is ascribed to the presence of a circumnuclear-obscuring torus \citep{antonucci84}. Hence, BLOs and HEGs can be considered as being part of the same class of objects, just seen at a different viewing angle. 

The study of the environment of various radio-loud AGN is intimately linked to both the morphology and the characteristics of the SMBH activity (such as the accretion rate, radio luminosity, and jet power), which is becoming fundamental in understanding whether or not there is a mutual impact between them (e.g., \citealt{gilmour07}, \citealt{ineson15}).
Various studies have already been conducted: all previous works concur with the conclusion that RGs prefer to inhabit large-scale galaxy-rich environments (e.g., \citealt{best04}, \citealt{tasse08}, \citealt{massaro20}) and for this reason they are often associated with galaxy clusters (e.g., \citealt{mcnamara05}, \citealt{giacintucci09}), where they play a significant role in regulating the
cooling flow of the intracluster medium (ICM) (e.g., \citealt{boehringer93}, \citealt{blanton03}, \citealt{mcnamara07}, \citealt{mcnamara12}). According to this scenario, RGs could be used as beacons of rich environments and ideal laboratories to test the cosmological scenario through the investigation of their formation and coevolution with the large-scale structure they inhabit (e.g., \citealt{bahcall72}).

However, early results highlighted differences in large-scale properties of FR~Is with respect to FR~IIs which tend to be found in galaxy-rich and isolated environments, respectively (e.g., \citealt{zirbel97}). A similar result was obtained by \citet{croston19} by cross-correlating optical catalogs of groups and clusters with RGs; they found a systematic correlation between the radio morphology of RGs and the richness of their environment, with a preferential pattern for FR I galaxies to inhabit richer environments than FR IIs. This would be in agreement with the scenario in which the morphological dichotomy depends on the density of the surrounding medium, where FR I jets are disrupted by an impact with a large-scale denser ambient (e.g., \citealt{jones79}, \citealt{blanton00}). \citeauthor{croston19} also discovered a correlation between cluster richness and radio luminosity. However, their sample is mostly formed by relatively low-luminosity RGs. \citet{ching17} found an analogous relation of galaxies' density with radio luminosity; they also showed that high-luminosity LEGs lie in a denser environment than HEGs and low-luminosity LEGs. A difference in the environment among the LEGs based on their luminosity and morphology was also found by \citet{capetti20}: the compact (and low-luminosity) FR~0s are usually  found in poor groups, while FR~Is  more often inhabit clusters of galaxies. 

In contrast, \citet{massaro20} found that radio sources in the local Universe, independently of both their morphological and optical classification, live in environments with a similar richness and characteristics, although their analysis included only a limited number of HEGs. Similarly, \citet{vardoulaki21} have found, by studying sources in the COSMOS field, that different types of environments are covered independently of the radio classification.

\begin{table*}[h!]
\caption{3C sources considered ordered for increasing redshift. The number of sources in the RS (N$_{RS}$) for RGs at $z > 0.05$ is the one found with the first method, without selecting just the extended sources.}

\label{tabb}
\begin{tabular}{l l r r r l r | l l r r r l r}
\hline
     3CR         &  z             & FR & Class            & A$_\text{v}$       & {\tiny{SDSS}} &  N$_\text{RS}$&    3CR         &  z             & FR & Class            & A$_\text{v}$       & {\tiny{SDSS}} & N$_\text{RS}$\\
\hline  
\color{red}{129} & 0.0208 & 1 & - &3.591 &      & 10$\pm$8  & 135      &  0.1253   &  2   & HEG           & 0.373  &     & 27      $\pm$     11  \\
066B & 0.0215 & 1 & LEG &0.304 &                 &  6 $\pm$ 20 & 424      &  0.127      &  2   & LEG              & 0.319  &                                                      & 75    $\pm$      21 \\ 
264 & 0.0217 & 1 &- & 0.078&                     & 17 $\pm$ 5  & 197.1   &  0.1301   &  2   & BLO                 & 0.139  & \checkmark                   &  31    $\pm$       13 \\ 
\color{red}{129.1} & 0.0222 & 1 & - & 3.789&    &  12 $\pm$ 7 &  223        &  0.1368   &  2   & HEG         &  0.038 & \checkmark                        & 35     $\pm$      9  \\ 
075N & 0.0232 & 1 & LEG & 0.6 &                  &  45 $\pm$ 8 & \color{red}{089}      &  0.1386   &  1   & - &  0.415 & \checkmark                        &  94    $\pm$       15 \\ 
402 & 0.0239 & 1 & - & 0.399  &                  &  22 $\pm$ 11 &  303  &       0.141     &  2  & BLO              &  0.063 & \checkmark                   & 45     $\pm$      14   \\ 
296 & 0.0240 & 1 & LEG &0.084 & &         0 $\pm$ 21       &   332        &  0.1517    &  2 & BLO                 & 0.077  & \checkmark                                   & 131    $\pm$       17  \\ 
083.1 & 0.0255 & 1 & - & 0.534& &       26 $\pm$ 10         &   348      &  0.154      &  1   & ELEG      &  0.314 &                                              & 113     $\pm$      19   \\ 
\color{red}{442} & 0.0263 & - & LEG &0.213 &     &  24 $\pm$ 9 & 273      &  0.1583   &   -   & BLO               &  0.069 & \checkmark                           & 38    $\pm$       8  \\ 
078 & 0.0286 & 1 & LEG & 0.565&                  &  10 $\pm$ 10 & 381      &  0.1605    &  2 & HEG        & 0.174  &                                                       & 116      $\pm$     32 \\ 
088 & 0.0302 & 2 & LEG &0.417 &                  & -4 $\pm$ 4  &   \color{red}{346}      &  0.161      &2 & -  & 0.224  & \checkmark  &  57 $\pm$ 14\\
338 & 0.0303 & 1 & LEG &0.037 &                  & 79 $\pm$ 10  &  357      &  0.1662    &  2 & LEG       & 0.153  & \checkmark                                   & 123    $\pm$       34 \\  
465 & 0.0303 & 1 & LEG &0.234 &                  & 87 $\pm$ 14  &   020      &  0.174     &  2   & HEG            &  1.323 &                                              &45         $\pm$  15   \\   
353 & 0.0304 & 2 & LEG &1.475 &                  &  39 $\pm$ 12 & 219      &  0.1744   &  2   & BLO               & 0.061  & \checkmark                           & 49    $\pm$       9 \\  
098 & 0.0304 & 2 & HEG & 0.741&                  &  -4 $\pm$ 6 &  063      &  0.175     &  2     & HEG    &  0.086 &                                                      &30     $\pm$      16  \\ 
076.1 & 0.0324 & 1 & - & 0.451&                   & 7 $\pm$ 8  & \color{blue}{033.1}   &  0.1809   &  2   & BLO       &  2.007 &        & 32 $\pm$ 16 \\
\color{red}{317} & 0.0345 & - & LEG & 0.123&     & 50 $\pm$ 14  &   061.1   &  0.184     &  2   & HEG              &  0.570 &                                              &  18    $\pm$       11\\ 
305 & 0.0416 & 2 & HEG &0.08 &                   & 10 $\pm$ 5  & 234      &  0.1848   &  2   & HEG                &  0.062 & \checkmark                           & 61     $\pm$      10 \\ 
029 & 0.0448 & 1 & LEG &0.117 &                  &  19 $\pm$ 13 & 018      &  0.188     &  2   & BLO              &  0.535 &                                     & 7         $\pm$  5  \\ 
\color{red}{293} & 0.0450 & - & LEG & 0.055&     &  4 $\pm$ 10 & \color{red}{319}      &  0.192     & 2 &   -  & 0.041  & \checkmark                       & 45 $\pm$ 10 \\ 
\color{red}{318.1} & 0.0453 & - & - &0.118 &     &  49 $\pm$ 11 &  213.1   &  0.194     &  2   & LEG              & 0.099  & \checkmark                  & 43    $\pm$       13 \\ 
\color{blue}{111} & 0.0485 & 2 & BLO & 5.458&    & 21 $\pm$ 5  & 028      &  0.1952   &  2   & ELEG       &  0.193 & \checkmark                           & 87      $\pm$     18 \\ 
\color{red}{371} &  0.0500   & -   & LEG & 0.112 &   &  (127 $\pm$ 32) 34     $\pm$      14  & 196.1   &  0.198     &  2   & LEG           &  0.216 &                                        &   28   $\pm$        9\\ 
 310 & 0.0535 & 2  & LEG  &  0.135 &  \checkmark & (70 $\pm$ 15) 25     $\pm$      12  & 401   &  0.2010     &   2  &         LEG      &  0.193 &                                           & 48 $\pm$ 19 \\ 
430 &  0.0541 & 2 & LEG & 2.035 &  & (353 $\pm$ 146) 40 $\pm$ 7  & 349   &  0.205     &  2   &   LEG      &  0.105 &  \checkmark                                  & 46 $\pm$ 11 \\ 
403.1   &  0.055     &  2     & LEG     &  0.760 &  &  (448 $\pm$ 99) 18   $\pm$       9 & 132   &  0.214     &  2   &   LEG      &  1.632 &                                              & 39     $\pm$      33 \\ 
390.3   &  0.0561   &  2   & BLO        &  0.236 &  &  (240 $\pm$ 39) 66     $\pm$      11  &  436   &  0.2145     &  2   &      HEG      &  0.298 &    \checkmark                          & 20 $\pm$ 9  \\ 
445      &  0.0562   &  2   & BLO       &  0.275 &  & (47 $\pm$ 14) 12     $\pm$      11  &  287.1   &  0.2159     &  2   &      BLO      &  0.082 &   \checkmark                           & 47 $\pm$ 11 \\ 
382      &  0.0578   &  2   & BLO       &  0.231 &  & (50 $\pm$ 29) 0   $\pm$       7  & \color{blue}{123}   &  0.2177     &  -   & LEG        &  3.233 &        & 19 $\pm$ 8 \\ 
403      &  0.0590   &  2   & HEG       &  0.626 &  & (368 $\pm$ 181) 6  $\pm$        6  &  017   &  0.2198     &  2   &         BLO      &  0.077 &  \checkmark                           & 38      $\pm$     9  \\ 
033      &  0.0596   &  2   & HEG       &  0.071 &  &  (72 $\pm$ 13) 34   $\pm$        9 & 459   &  0.2199     &  2   &  BLO      &  0.216 &                                   & 19 $\pm$ 7 \\ 
192 & 0.0598 & 2 & HEG&  0.183 & \checkmark      & (22 $\pm $ 14) -2   $\pm$        4  & 180   &  0.2200     &   2  &        HEG      &  0.325 &                                           &  12 $\pm$ 12\\ 
136.1 & 0.064 & 2 & HEG & 2.494 &  & (135 $\pm$ 34) 9 $\pm$ 3  & 456   &  0.2330     &  2   &    HEG      &  0.123 &                &  8 $\pm$ 7 \\ 
\color{red}{035}    &  0.0670 & 2  & - &  0.461 &  & (122 $\pm$ 50) 15   $\pm$        5 & 171 & 0.2384 & & HEG & & & 9 $\pm$ 7 \\
015  &  0.073  & 1   & LEG &  0.071 & \checkmark &  (29 $\pm$ 14) 10   $\pm$        9 & 284   &  0.2394     &  2   &         HEG      &  0.052 &   \checkmark   &  24 $\pm$ 8 \\ 
285  &  0.0794   &  2 & HEG &  0.056 & \checkmark&  (40 $\pm $ 15) 24   $\pm$        7 &  093.1   &  0.2430     &  -   &      HEG      &  1.283 &                                                &   27    $\pm$       9\\ 
452  &  0.0811   &  2   & HEG &  0.459 &         & (84 $\pm$ 30) 33     $\pm$      10 & 166   &  0.2449     &    2 &  LEG      &  0.673 &                & 23 $\pm$ 12 \\ 
198  &  0.0815  & 2& SF &  0.086 & \checkmark    &  (72 $\pm$ 20) 37   $\pm$        7 & 288   &  0.246     &   1  &  LEG      &  0.026 &   \checkmark   &  22 $\pm$ 12 \\
277.3 &  0.0857 & 2 & HEG & 0.038 & \checkmark   & (46 $\pm$ 24) 24   $\pm$        14 &  410   &  0.249     &   2  &        BLO      &  4.521 &        &  25 $\pm$ 10\\
227  &  0.0861 & 2  & BLO &  0.085 & \checkmark  & (46 $\pm$ 10) 38   $\pm$        7 & 079   &  0.2559     &  2   &         HEG      &  0.424 &                                         &  7       $\pm$    6  \\ 
105  &  0.089     &  2   & HEG &  1.588 &        & (34 $\pm$ 11) 25   $\pm$        6  & 379.1   &  0.256     &  2   &       HEG      &  0.201 &                & 18 $\pm$ 10 \\ 
326  &  0.0895 & 2 & LEG &  0.170  &  \checkmark &  (54 $\pm$ 13) 40    $\pm$       9 & 323.1   &  0.264     &   -  &         BLO      &  0.141 &    \checkmark  & 21 $\pm$ 7 \\
388  &  0.091     &  2   & LEG &  0.260 &        & (171 $\pm$ 25) 80       $\pm$    10   & 303.1   &  0.267     &  2   &         HEG      &  0.120 &                & 10 $\pm$ 6 \\ 
321  &  0.096  &  2 & LEG &  0.145 &  \checkmark & (27 $\pm$ 9)  13       $\pm$    5  & 460   &  0.268     &  2   &      LEG      &  0.297 &                & 12 $\pm$ 6 \\ 
236  &  0.1005   & 2 & LEG & 0.036 &  \checkmark &  25 $\pm$ 10 & 300   &  0.27     &  2   &      HEG      &  0.116 &  \checkmark    & 27 $\pm$ 11 \\ 
433  &  0.1016    &  1-2   & HEG&  0.483 &       &  109     $\pm$      37  &   153   &  0.2769     &   2  &       LEG      &  0.537 &                                                & 50    $\pm$       12  \\ 
327  &  0.1041    &  2 & HEG    &  0.295 &          & 74    $\pm$       16  & \color{blue}{133}   &  0.2775     &  2   &     HEG &  2.967 &        & 21 $\pm$ 9 \\ 
223.1 & 0.107 & 2 & HEG & 0.058 & \checkmark     &  30   $\pm$        10  &  \color{red}{052}   &  0.2854     &  2   & -  &  0.789 &        & 28 $\pm$ 10 \\ 
315    & 0.1083 & - & LEG &  0.205    & \checkmark &  67    $\pm$       16  &    458   &  0.289     &   2  &       HEG      &  0.277 &                & 12 $\pm$ 9 \\
\color{red}{130}    &  0.1090  &  1 & - & 4.535  & & 56 $\pm$ 17  &  \textcolor{red}{438}   &  0.290     &    1 &         -        &  1.221 &                & 109 $\pm$ 20 \\   
184.1   &  0.1182   &  2   & BLO         & 0.108  & &  23     $\pm$      10   &  173.1 &  0.2921 & & LEG & & & 8 $\pm$ 9\\
314.1   &  0.1197    &   -  & ELEG               & 0.063  &                                               &      45     $\pm$      9 & 165   &  0.2957     &  2   &    LEG      &  0.572 &                & 21 $\pm$ 12  \\

\hline
     \end{tabular}
   
     \medskip
Column description: (1) Source name; (2) Redshift; (3) FR classification; (4) Optical spectroscopic classification; (5) Galactic absorption in magnitudes; (6) Availability of SDSS data; (7) Number of RS sources subtracted to the ground. Values outside the parenthesis  refer to the estimates of $N_{RS}$ obtained by selecting only extended optical sources for the sources with $0.05<z<0.1$.  

\end{table*}

In this work we investigate the environment of RGs listed in the Third Cambridge Catalog of radio sources (3CR) (\citealt{bennett62b,bennett62a}) up to $z = 0.3$. 
These criteria predominantly select FR~II RGs, as the low-luminosity FR~Is are more common at low redshifts (i.e., $z<0.1$) and include a larger number of HEGs with respect to the sample studied in \citet{massaro19}, since HEGs are exclusively FR~II galaxies, as mentioned above.
This 3CR subsample has radio luminosities spanning over four orders of magnitude, and with an almost complete spectroscopic classification from \citet{buttiglione10}, allowing us to perform the analysis between various classes.
In order to make an environmental comparison between different subclasses, it is important to consider both a morphological and spectroscopic classification. The 3CR sample has all of this information available; moreover, the sky area covered by the 3CR sample is wide enough to include a large number of bright sources, contrary to other larger catalogs. 

Our analysis will provide robust statistical results when large-scale environments of both optical spectroscopic classes and morphological types are analyzed and compared.
In particular, HEGs and LEGs can be used to investigate whether the
level of activity of the SMBH and the large-scale properties are correlated,
while FR Is and FR IIs can be employed to search for a correlation between the environment as well as the
radio morphology and luminosity. Furthermore, we can also test the prediction of the unified model that postulates that no differences should be found between the environment of HEGs and BLOs.
Finally, as several of the  works described above were also conducted with a subsample of 3C sources, we can compare our conclusions to the ones found in previous analyses.

We characterized the environment of RGs by studying their color-magnitude diagrams
(CMDs) and counting the number of galaxies located in the so-called red
sequence (RS).
The RS is a tight relation between the color and magnitude of galaxies belonging to a group or cluster of galaxies. It was discovered thanks to an observation of elliptical galaxies (e.g., \citealt{visvanathan77}).
RSs are observed in galaxy clusters up to $z \sim 2$ (e.g., \citealt{gobat11}) and can be exploited in the search for large-scale structures with photometric data.
The RS is described by the following three parameters: a zero point, slope, and photometric dispersion. These parameters are different depending on the redshift considered since the properties of the RS vary with the local galaxy density (\citealt{valentinuzzi11}). In particular, the RS dispersion is larger in lower density regions (i.e., in galaxy groups with respect to 
clusters). At a lower redshift, galaxies' groups are more abundant than clusters, leading to an increased RS dispersion with respect to more distant regions.
However, as shown by \citet{mei09}, the rest-frame zero point of the RS 
shows no significant evolution out to redshift z $\sim$ 1, indicating that the RS was already in place $\sim$6 billion years ago.

This paper is organized as follows: in Sect. \ref{observations} we describe the sample
of the selected sources and the available optical observations with the Panoramic Survey Telescope \& Rapid Response System (Pan-STARRS) (e.g., \citealt{chambers16}) \footnote{Web page:https://outerspace.stsci.edu/display/PANSTARRS/} and Sloan Digital Sky Survey (SDSS; e.g., \citealt{albareti17})\footnote{Web page: https://www.sdss.org}. In Sect. \ref{metodo} we discuss the method we applied to obtain the richness
of each RG environment. The results emerging from the comparison of optical and morphological classes are reported and discussed in Sect. \ref{results}. 
A summary is found in Sect. \ref{summary} where we also draw our
conclusions. We adopted the following set of cosmological parameters: H$_{0} = 69.7$ km s$^{-1}$ Mpc$^{-1}$ and $\Omega_{m} =0.286$ (\citealt{bennett14}).

\section{The 3CR sample and adopted data}
\label{observations}

We initially considered the 3CR catalog of \citet{bennett62b,bennett62a}.
The original sample comprises 328 radio sources and \citet{spinrad85} were able to identify the optical counterpart and redshift of 298 objects among them. Starting from these 298 radio sources, we examined the subsample that formed by 104 RGs at redshift $0.02 < z < 0.3$ for which we have the spectroscopic classification from \citet{buttiglione09}. Their main parameters are listed in Table~\ref{tabb}.

Nine objects and five objects (highlighted in red in
Table~\ref{tabb}) do not have the optical or morphological
classification, respectively, and they are not included in the analysis. Furthermore, poorly represented groups of sources (the single star-forming galaxy 3CR~198 and the three extreme low excitation galaxies 3CR~314.1, 3CR~348, and 3CR~028) are not used
in our statistical study. Finally, four sources (in blue in
Table~\ref{tabb}) are characterized by either a combination of high Galactic absorption ($A_v > 2$ magnitudes) and relatively high redshift, or just a very high absorption ($A_v > 5$ magnitudes). As a result, our analysis could not be performed at the requested level of absolute magnitude (see Sect. 3 for more details) and these sources have also been excluded. 

The analysis was performed separately in two redshift ranges: the morphological types (i.e., FR I and FR II) were compared to each other in the $0.02 < z < 0.1$ redshift bin, while the optical spectroscopic classes (i.e., HEG, LEG, and BLO) were compared in the $0.05 < z < 0.3$ bin.
In the former case, the lower redshifts were necessary to include FR~I RGs in the sample, but the background contamination is significantly higher (see Sect.~\ref{parte2}). In addition, the methods used to compare the different classes in the two ranges (also described in Sect.~\ref{parte2}) are different.
Hence, dividing our RGs into different redshift bins is important not only to provide us with a complete uniform sample of sources, but also to allow the same method to be applied to the data consistently.

The final samples include 23 LEGs, 29 HEGs and 16 BLO RGs, and 14 FR Is and 24 FR IIs. In these two subsamples, a total of 14 and six sources, respectively, have been excluded because of the reasons stated above.

\section{Method}
\label{metodo}

We used the optical photometric catalog of Pan-STARRS, with its median  point spread function (PSF) of 0.94'', which produced images of the whole sky north of $-30^{\circ}$ declination in five broadband filters ($g, r, i, z$, and $y$) over multiple epochs. This
allowed us to produce stacked images with
an average coverage of $\sim 8.9$ exposures per pixel, reaching apparent magnitude limits of 23.3, 23.2, and 23.1 in the $g$, $r$, and $i$ band, respectively.

\begin{figure*}
\includegraphics[width=0.49\textwidth]{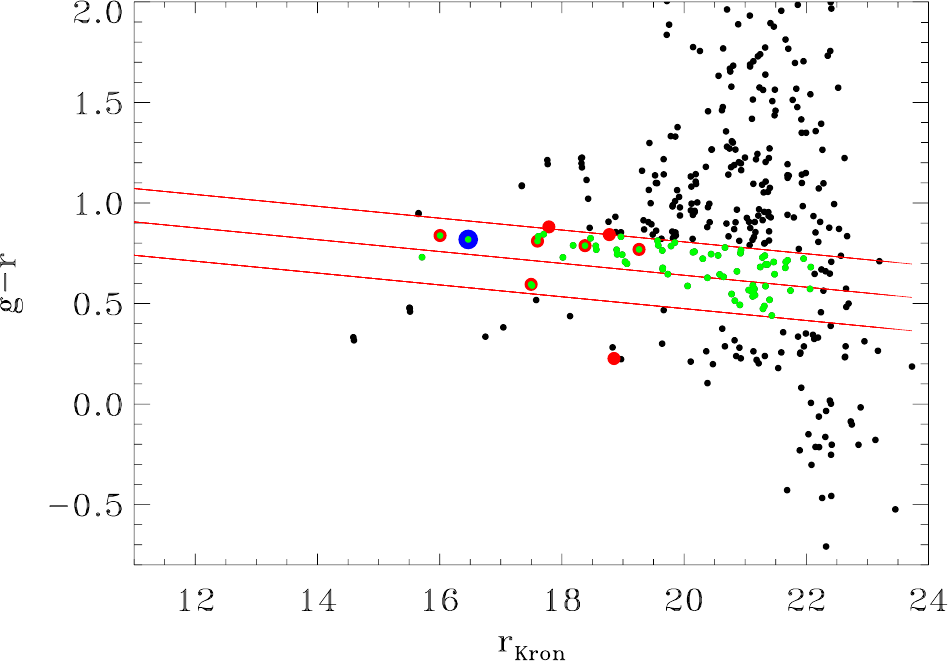}
\includegraphics[width=0.49\textwidth]{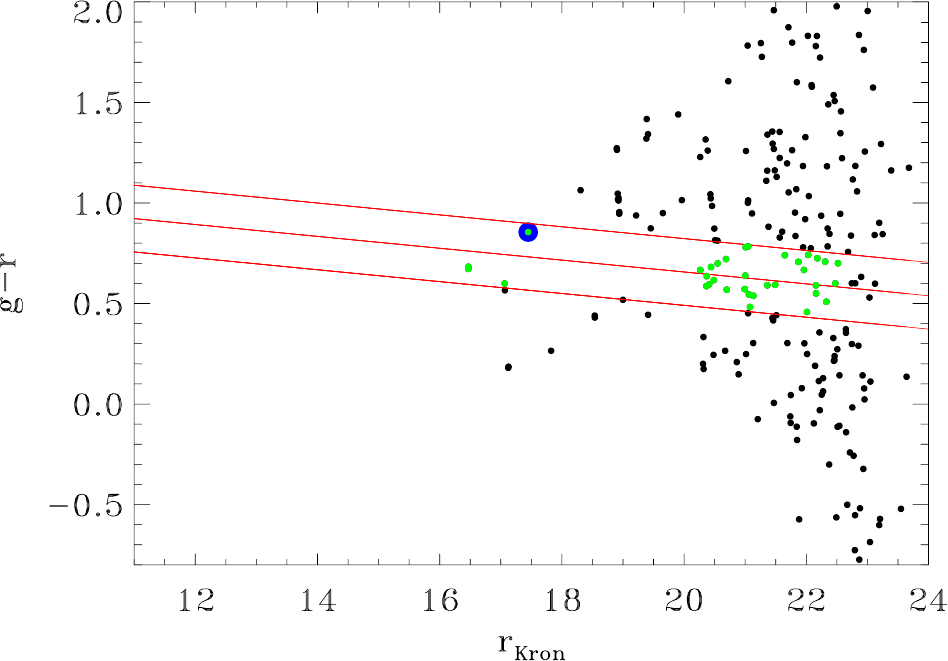}
\caption{{\it{Left panel:}} CMD of the source
  3CR~089. The blue dot represents the host of 3C089; the red dots are
  the spectroscopic companions identified through SDSS data; the red
  lines represent the boundaries of the RS from \citet{omill19}; and
  the green dots are the sources falling inside the RS with absolute
  magnitude $M_r < -17$. {\it{Right panel:}} CMD
  of the source 3CR~063. The RS in this source is less clearly defined. }
\label{Fig1}
\end{figure*}

We searched for the photometric data of all sources inside a circular
region with a radius of $r = 500$ kpc centered on each 3CR object. In addition, four background regions of the same size were chosen at a projected distance of 5 Mpc from each source in the north, south, east, and west directions.\footnote{All physical distances were computed at the redshift of the central RG.} The final backgrounds, used throughout the analysis, have been defined for each RG environment as the average of the four background fields taken into account.

As mentioned in the Introduction, galaxies belonging to the same galaxy group or cluster align along an oblique stripe in the CMD, defining a red sequence. The position of the stripe, its thickness, and inclination depend on the group or cluster redshift and mass density.
Our work is based on defining the richness of each 3CR RG environment by measuring the excess of sources included in the RS with respect to the background.
We adopted the RS relations found by
\citet{omill19} that are defined in the $g - r$ color
calculated with the SDSS filters in the following three
redshift bins: $z \leq 0.065$, $0.065 < z < 0.1$, and $0.1 < z < 0.2$. 

To apply the results of \citet{omill19} based on the SDSS filters to the Pan-STARRS photometric system, we calculated the photometric correction to the RS parameters
using the coefficients reported in \citet{tonry12}.

To extend the study to galaxies up to $z = 0.3$, outside the range covered by the \citeauthor{omill19} analysis, we extrapolated the RS parameters, 
calculating the rest-frame magnitudes and colors
by applying the K correction from
\citet{chilingarian10} to our photometric data.

\begin{table}
\caption{RS parameters from \citet{omill19} listed for increasing redshift bins. The last row in the z=0.2$-$0.3 bin reports the zero-point value calculated by our RS extrapolation and the other parameters adopted for the main analysis.}
\begin{tabular}{r r c r}
\hline
Slope & Zero-point & Phot. Disp. & z range     \\
     \hline
 -0.032$\pm$0.006 & 0.269$\pm$0.132 & 0.312 & < 0.065\\
 -0.035$\pm$0.007 & 0.111$\pm$0.166 & 0.093 & 0.065 $-$ 0.1\\
 -0.034$\pm$0.019 & 0.103$\pm$0.341 & 0.166 & 0.1 $-$ 0.2\\
 -0.034 & 0.04 & 0.312 & 0.2 $-$ 0.3\\
\hline   
\end{tabular}

     \label{tab7} 
     \end{table}

\subsection{Comparison of HEG, LEG, and BLO at $0.05<z<0.3$}\label{comp1}

For each source in the redshift range $0.05<z<0.3$, we produced a CMD of the optical sources within a radius of 500 kpc, adopting the aperture magnitude color indexes $g - r$. Kron photometry was used for an appropriate estimate of the extended sources' magnitude in the $r$ band. Moreover, we corrected colors
for the Galactic reddening using the \citet{fitzpatrick99} extinction law.

Fig.~\ref{Fig1} shows two examples of the CMD.
The filter-corrected RS relations extrapolated by \citet{omill19} are drawn in red in each CMD.
In the left panel, we report the
case of 3CR~089, where a RS is clearly seen. 
To further highlight the RS in the diagrams, we also considered (when available, see Tab. \ref{tabb}) the
spectroscopic SDSS data. These data can be exploited to search for
companions (i.e., objects at the same redshift) that are marked in the respective CMD. The position of these spectroscopic companions into the RS generally confirms the goodness of the relations of  \citet{omill19}. Only one of the companions lies below the lower limit of the RS dispersion, at $g-r \sim 0.25$, which is likely a star-forming galaxy. Star-forming galaxies have been revealed by previous studies which highlighted the presence of a blue cloud region in the CMDs (e.g., \citealt{eales18}). In this case, the $g-r$ color index of this particular source in the field of 3CR~089 corresponds to the typical values of late-type galaxies found in the blue clouds of \citet{eales18}.

The right panel instead shows the CMD for
3CR~063 where the RS is not readily visible and no SDSS data are available.

\begin{figure}
\includegraphics[width=0.49\textwidth]{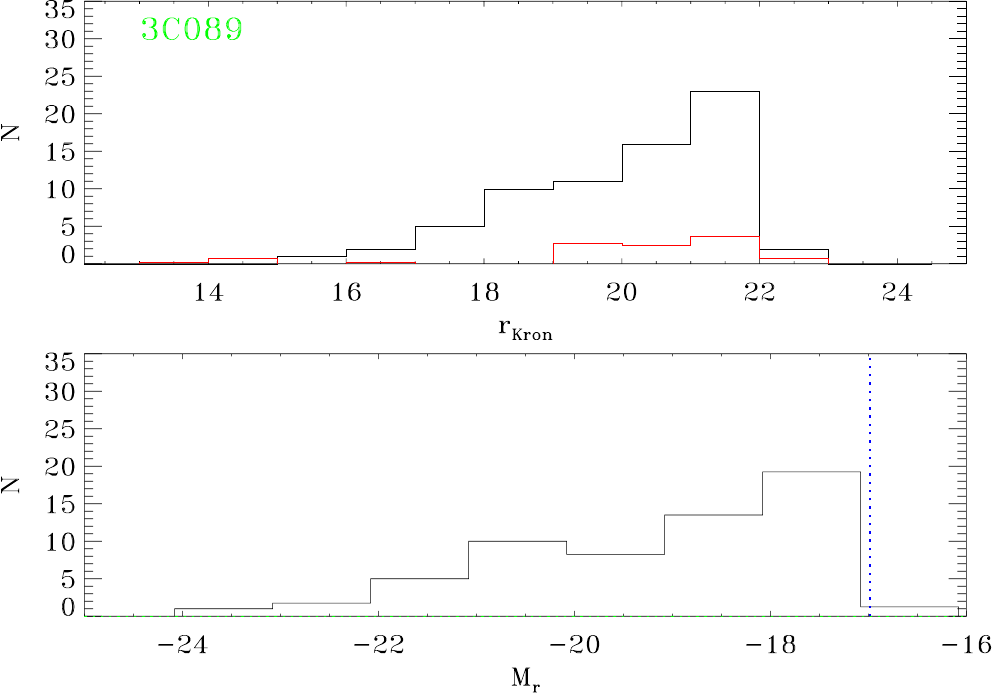}
\caption{{\it{Upper panel}}: Histogram showing the number of sources in the various magnitude bins falling into the RS (black line) and the number of sources of the corresponding background RS (red line). {\it{Lower panel}}: Histogram of the number of sources in the RS after background subtraction. The vertical blue dotted line marks the magnitude limit adopted.}
\label{Fig2}
\end{figure}

Fig.~\ref{Fig2} shows the distribution of the relative magnitude of the sources within the RS for 3CR~098 (black
histogram) compared to the averaged value in the four background fields (red histogram). In the lower panel, the background was subtracted to estimate the excess of sources for which we report the K-corrected absolute magnitude in the $r$ band.  

We estimated the excess of sources for each 3CR RG, a parameter directly related to the local galaxies' density. In order to properly compare the results of sources at different redshifts, we adopted a fixed cut in absolute magnitude. Moreover, the
comparison between sources of different redshifts could be done only if we considered the same RS area from which the number of sources was
extracted. Hence, the same RS dispersion was adopted.

The results reported in Table \ref{tabb} refer to a RS dispersion equal to 0.312 (appropriate for the lowest redshift sources) and to a threshold in K-corrected absolute magnitude of $M_r = -17$.  The selected galaxies are $\sim$3.5 magnitudes fainter than  the characteristic luminosity L$^*$ of the luminosity
  function of local early-type galaxies (see,
  e.g., \citealt{bell03b}).
The uncertainties on N$_{RS}$ were calculated by adding, in quadrature, the error on the background fields to the Poissonian noise on the source field.
In particular, for each source the background error has been defined as the standard deviation of the four background fields considered for that source. The Poissonian statistic on the background does indeed underestimate the error because it does not take any other effects into account, such as cosmological variations or the possibility of including another group or cluster of galaxies  by chance in one of our background fields.

\subsection{Comparison of FR Is and FR IIs: A different approach at $0.02 < z < 0.1$}\label{parte2}
  
\begin{figure}
\includegraphics[width=0.45\textwidth]{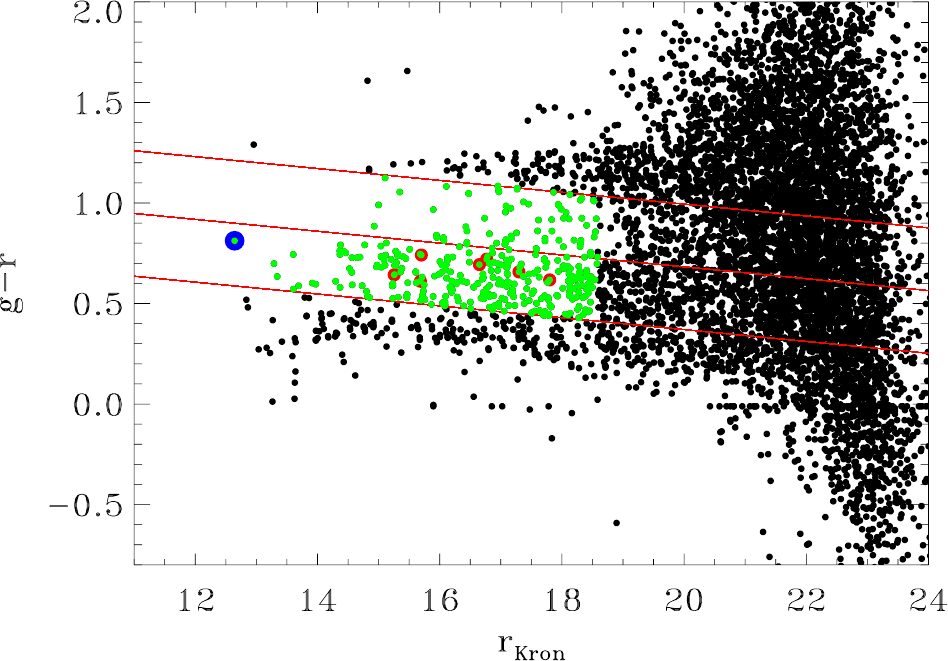}
\caption{CMD of the source 3C~338. The blue dot represents the RG and the green ones are all the sources falling into the RS relation. The red dots are the spectroscopic members identified with SDSS.}
  \label{3C338rgr}
\end{figure}

For the comparison of the environment of the FR~Is and FR~IIs, we extended the analysis to a lower redshift threshold in order to include a sufficient number of FR~I sources. We then considered the $0.02 < z < 0.1$ redshift range. However, this has a strong
impact on the CMDs because the number of sources included increases dramatically, as shown in Fig.~\ref{3C338rgr}, where the CMD of the RG 3CR~338 (at z=0.0303)
is shown. This is due to the fact that the
area covered by the 500 kpc radius becomes as large as 20$^\prime$ for a source at $z=0.02$. As a result, also the number of sources in the background fields, effectively setting the uncertainty on the measurement of $N_{\rm RS}$, increases, compromising our ability to estimate the local galaxies' density.
We then adopted a different approach than the one used to compare sources at a higher redshift in the attempt to reduce the number of spurious sources.

\begin{figure}
\includegraphics[width=0.45\textwidth]{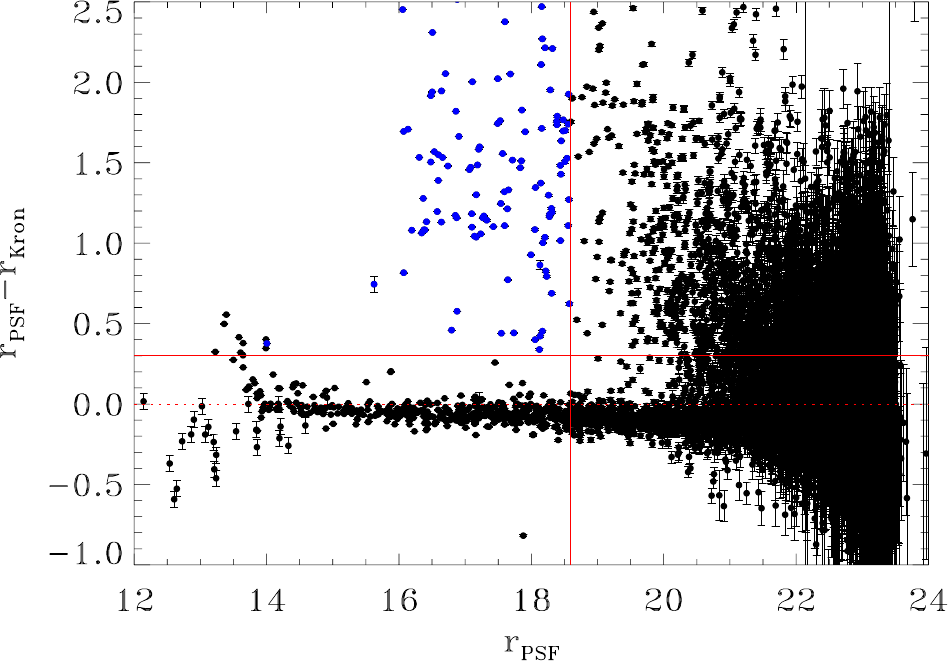}
\caption{Comparison of the PSF and Kron magnitudes. Their
  difference is  approximately zero for point sources (i.e., stars), while the fluxes of the extended galaxies are under-estimated by the PSF photometry. This creates a sharp separation between them. The two solid lines mark the boundaries of the regions where extended galaxies are located.}
\label{psfkron}
\end{figure}

\begin{figure}
\includegraphics[width=0.45\textwidth]{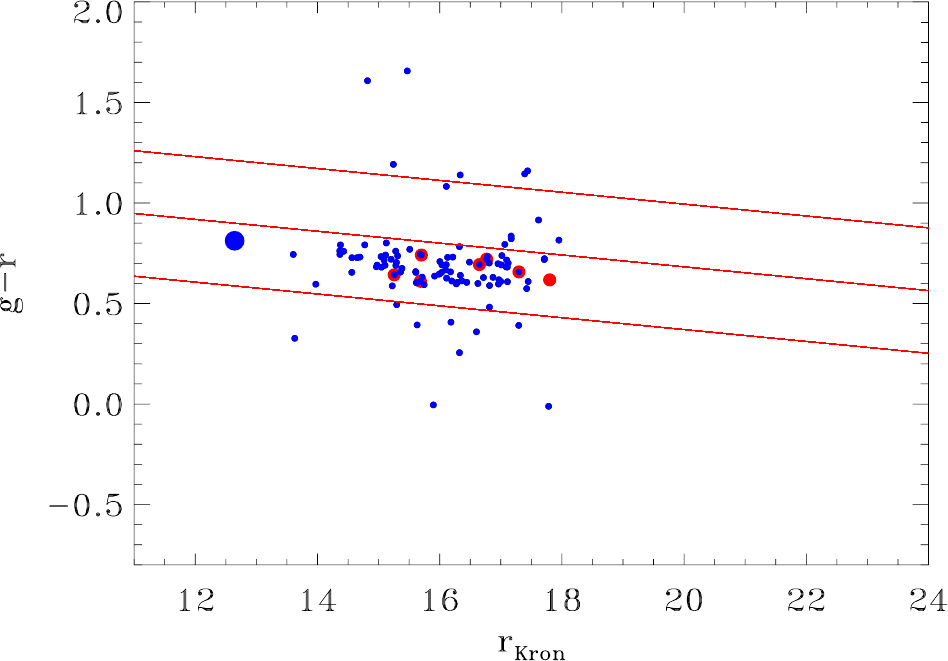}
\caption{CMD of the source 3C~338 but the point-like sources have   been removed. The remaining sources are all extended and form a well-defined RS.}
\label{3C338_2}
\end{figure}

In fact, at these low redshifts, it is possible  to separate extended sources (i.e., galaxies) from point-like
sources by comparing the PSF ($r_{PSF}$) and the Kron ($r_{k}$) magnitudes provided by the Pan-STARRS catalog. The difference in magnitude of stars is $r_{PSF} - r_{k} \sim 0$. \footnote{This difference is not exactly zero for stars, as Kron magnitudes require a correction to convert them into total magnitudes.}  On the other hand, galaxies have $r_{PSF} > r_{k}$.  Fig.~\ref{psfkron} shows
the distribution of the difference between these two magnitudes, that is, ($r_{PSF} - r_{k}$) versus $r_{PSF}$. A branch with $r_{PSF} - r_{k} \sim 0$ is readily visible, which formed by the point-like sources.\footnote{This branch shows a large spread for high fluxes, due to saturated stars.} A large number of sources instead show $r_{PSF} > r_{k} $ and these can be readily separated from the stars down to $r_{PSF} \lesssim 19$. 
The best separation between the two populations was obtained by adopting $r_{PSF} - r_{k} > 0.3$. To estimate the excess of sources, we used the same limit in absolute magnitude adopted for the RS, that is, $M_r < - 17$, but considering only extended sources. Fig.~\ref{3C338_2} shows the 
impact of selecting only the extended sources: the number of sources is drastically reduced and they form a well-defined RS. Similarly, the number of sources in the background fields, selected with the same method, decreases and thus reduces the uncertainty on $N_{\rm RS.}$.

Some low redshift sources (in the range $0.02 < z < 0.05$) presented a high uncertainty for $N_{\rm RS}$ associated with a significant fluctuation of the background values.
We noticed how this effect was systematically attributed to those sources with low Galactic latitudes, due to the large gradient in Galactic sources. For these RGs we considered four background regions all located at the same Galactic latitude. 

\section{Results}
\label{results}

\begin{table}
\caption{Results of the statistical tests on $N_{RS}$ distributions for LEGs, HEGs, and BLOs at 0.05 $<$ z $<$ 0.3 and FR Is and FR IIs at 0.02 $<$ z $<$ 0.1.}
\begin{tabular}{l l}
\hline
Sample &    K-S probability            \\
     \hline
HEG-FR~II vs. BLO-FR~II & 0.5 \\
HEG-FR~II vs. LEG-FR~II & 0.22 \\
HEG+BLO-FR~II vs. LEG-FR~II & 0.26 \\
FR I vs. FR II &  0.5 \\
\hline   
Sample & Median \\
\hline
BLO-FR~II & 45 $\pm$ 18 \\
HEG-FR~II & 30 $\pm$ 16 \\
LEG-FR~II & 48 $\pm$ 31\\
BLO+HEG-FR~II & 34 $\pm$ 12 \\
FR I & 19 $\pm$ 9 \\
FR II & 24 $\pm$ 5\\
\hline

    \hline
\end{tabular}

     \label{tab6} 
     \end{table}

\begin{figure*}
\includegraphics[width=0.45\textwidth]{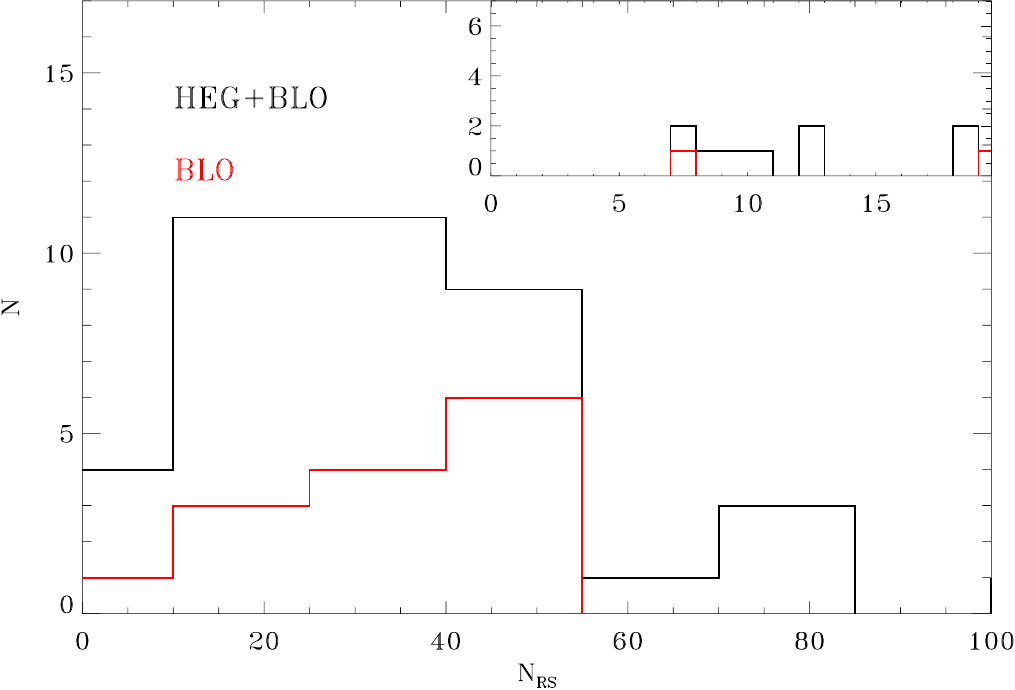}
\includegraphics[width=0.45\textwidth]{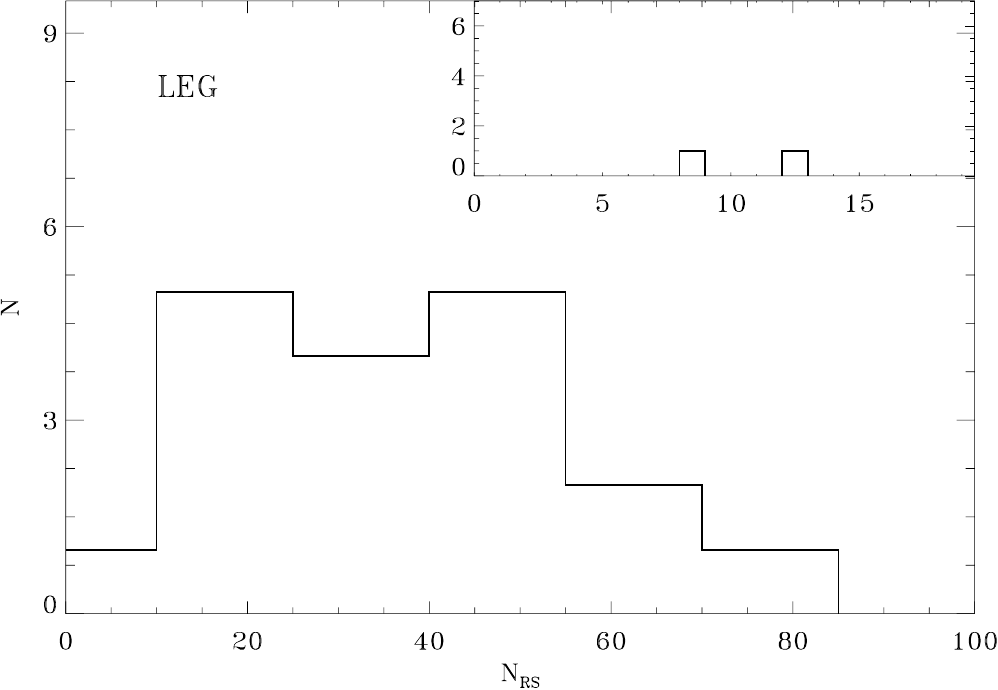}
\caption{Histograms showing the distribution of the number of RS
  objects for the HEGs and BLOs (left panel) and LEGs (right
  panel). The insets show an enlargement of the same distributions with a smaller bin
  size. The magnitude cut used is $M_r$ = -17, while the RS dispersion is
  0.312.}
\label{finalhist}
\end{figure*}

For all RGs  we estimated the total number of sources in the RS and we subtracted its respective average of the background fields. We estimated the uncertainty on the source count excess as described in Section~\ref{comp1}.

\subsection{Results for optical spectroscopic classes}

We started by considering the results for the sources with $0.05<z<0.3$, comparing the different optical spectroscopic classes. The results are reported in Tab. \ref{tab6} and are
visualized in the histogram of Fig.~\ref{finalhist}, where we plotted their richness (i.e., how many sources belong to their RS, $N_{RS}$) distribution.

Both histograms reveal a variety of environments inhabited by these
RGs, from the poorest one with a small excess of objects found in the RS area to the richest ones with more
than a hundred sources (i.e., rich galaxy clusters).  However, both LEGs and HEGs avoid isolated fields and prefer to be in small groups.
We conclude that
RGs can be found in both large-scale galaxy-rich
environments (i.e., galaxy clusters) and in poorer
fields, with no preferential patterns.

We then compared the distributions of $N_{RS}$ of the various classes with a Kolmogorov-Smirnov test. In the case of BLOs and HEGs, the two classes are not statistically different (see Table~\ref{tab6}), which is as expected
from the UM, although they are within the relatively small number
of BLOs. The same result  from the comparison of HEG/BLO and LEG types also
applies.
Furthermore, the values of the median of $N_{RS}$ for the various classes are consistent within errors.

We also tested the suggestion of the presence of a correlation between radio luminosity and $N_{RS}$. A Spearman rank test indicates that such a connection is not present for our sample as we obtained a probability of correlation $<$0.01.

The results presented above are based on the assumption of a specific set of values for the parameters describing the RS, that is, the zero point and dispersion of the RS, and a limit on the absolute magnitude. We then tested whether these values affect our conclusions. We then considered two different limits in absolute magnitude ($M_r=$-16 and $M_r$=-18) and varied the dispersion, adopting the different values derived by \citet{omill19} in turn in the various redshift ranges. 
Thus, we repeated the same analysis using the other two values for the dispersion and slope listed in Table~\ref{tab7}.
The environment richness depends on the RS parameters, as well as on the magnitude limit, but the final results are unchanged: the statistical tests on the distributions show no significant difference among the various classes. 
The adoption of the wider RS dispersion is preferable and gives a better estimate of the local galaxies' density. This is due to a general increase in the signal-to-noise ratio.
The signal-to-noise ratio was calculated as  $N_{RS} / \sigma$, where $\sigma$ is the uncertainty associated with $N_{RS}$.

\begin{figure}
\includegraphics[width=0.49\textwidth]{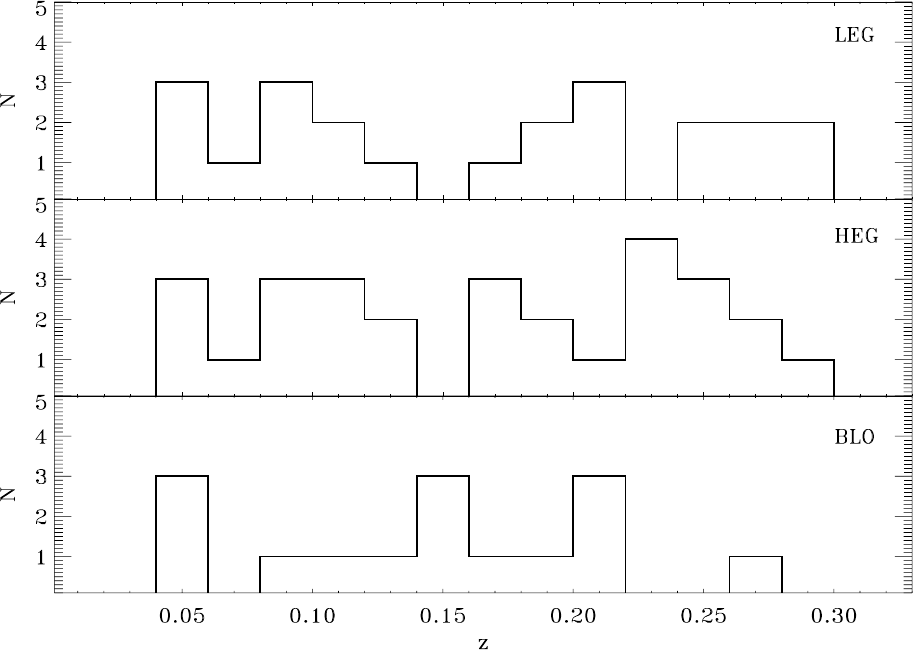}
\caption{Redshift distributions of LEGs ({\it{upper panel}}), HEGs ({\it{middle panel}}), and BLOs ({\it{lower panel}}) in the redshift range $0.05 < z < 0.3$. The statistical tests show that they are not statistically different for the three classes of RGs .}
\label{FigZ}
\end{figure}

Finally, although our analysis is aimed at limiting the effects of the different redshift of the sources considered as much as possible, we cannot exclude that some residual  effect is still present. However, the redshift distribution of the three classes, shown in Fig. \ref{FigZ}, is not statistically different as obtained from the Kolmogorov-Smirnov test and the median values (see Table~\ref{tab3} for a summary of the statistical comparison).

\begin{table}
\caption{Results of the statistical tests on the redshift distributions for LEG, HEG, and BLO.}
$$
\begin{tabular}{l l}
\toprule
Sample &    K-S probability            \\
     \hline
HEG vs. BLO & 0.32 \\
HEG vs. LEG & 0.96 \\
HEG+BLO vs. LEG & 0.86 \\
\hline
Sample & Median \\
\hline
BLO & 0.15 $\pm$ 0.02 \\
HEG & 0.18 $\pm$ 0.02 \\
LEG & 0.19 $\pm$ 0.02\\
BLO+HEG & 0.16 $\pm$ 0.01 \\
\hline

    \hline
\end{tabular}
$$
     
     \label{tab3} 
     \end{table}

\subsection{Results for morphological classes}

\begin{figure*}
\includegraphics[width=0.45\textwidth]{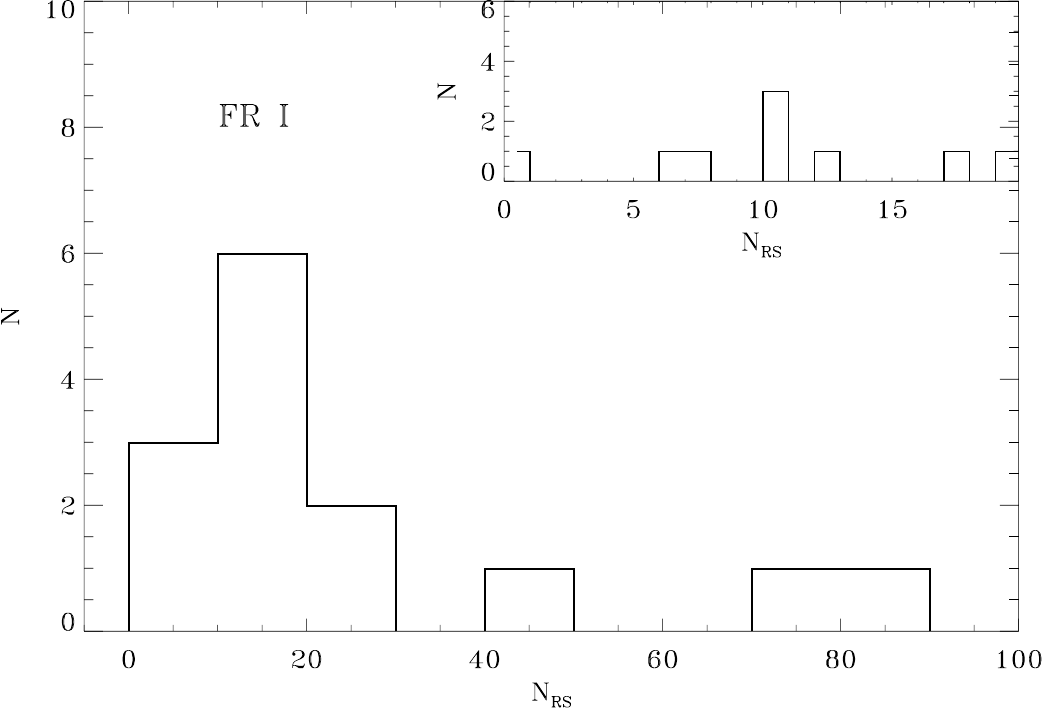}
\includegraphics[width=0.45\textwidth]{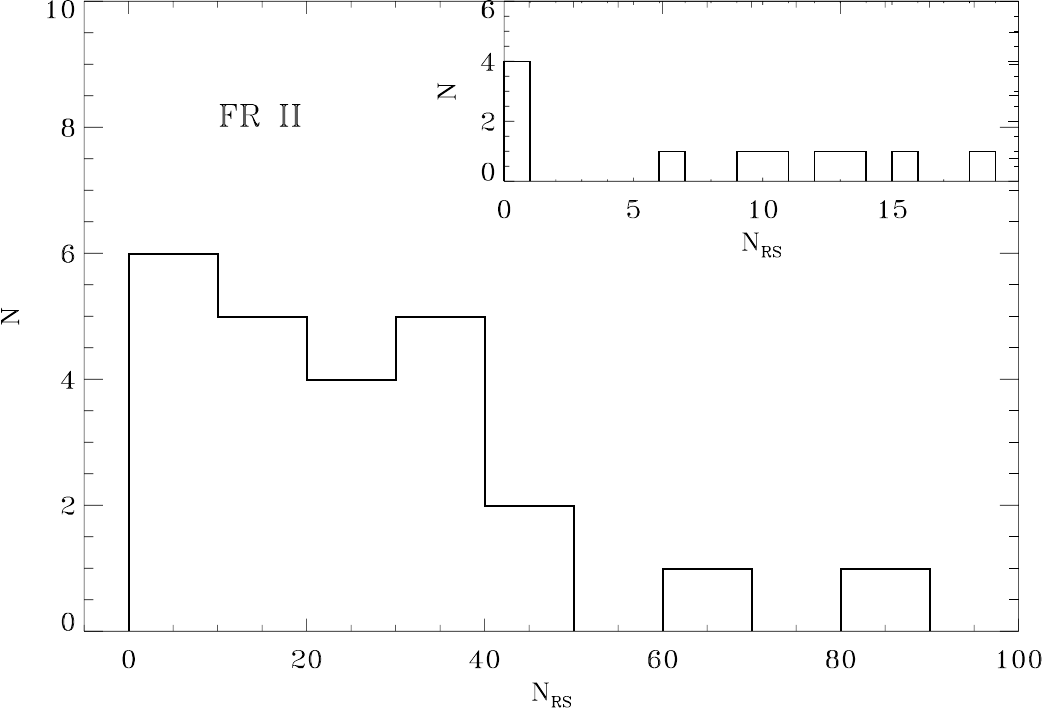}
\caption{Histograms showing the distribution of $N_{RS}$ for the FR~Is (left panel) and FR~IIs (right
  panel). The insets show an enlargement of the same distributions with a smaller bin
  size. The magnitude cut used is $M_r=$-17, while the RS dispersion is
  0.312}
  \label{finalhistFR}
\end{figure*}

We here analyze the results obtained from the comparison between FR Is and FR IIs, as visualized in the histograms of Fig.~\ref{finalhistFR} where we report their richness distribution. As for the results obtained for spectroscopic classes,
both histograms reveal a variety of environments inhabited by FR Is and FR IIs. We conclude that RGs can be found both in large-scale galaxy-rich environments and in isolated or poor
fields, independently of their radio morphological classification.
Both the Kolmogorov-Smirnov test and the median values on the FR I and FR II distributions (Table~\ref{tab6}) indicate that the two classes are not statistically different. This result is robust against changes in the different RS parameters adopted.

Similarly to the results obtained for the FR~IIs, no correlation between the radio luminosity and $N_{RS}$ is found.

\section{Summary and conclusions}
\label{summary}

The aim of this project was to explore the large-scale environment of RGs. We selected a sample of 104 3CR RGs with $0.02<z<0.3$, considering both their optical classification into LEGs, HEGs, and BLOs types, and morphological classification into FR~Is and FR~IIs. 

From the Pan-STARRS catalog, we extracted the objects located in a circular area with a radius of $r = 500$ kpc centered on each source, and we built $g - r$  versus $r$ CMDs.
Early-type galaxies belonging to a galaxy group or cluster are expected to lie along a RS in this diagram.
The richness of each 3CR source has been defined as the difference between the number of objects falling into the RS area of the source and that in the background, defined as the average of four different regions of the same area at a 5 Mpc distance from the source.

However, for the sources located at the lowest redshifts (i.e., $z<0.05$), the stellar contamination in the CMD becomes dominant and it substantially increases the uncertainties on the local galaxies' density. For these sources the analysis required a further selection of sources, that is, we considered only the extended ones.
We exploited Kron and PSF magnitude estimations to identify extended objects (i.e., galaxies) in the fields before the calculation of richness.
This method cannot be applied up to $z \gtrsim 0.1$, because the angular size of galaxies is not sufficient to provide a clear selection.

We obtained the following results:
\begin{itemize}
\item HEGs' and BLOs' richness distributions are not statistically different, which is as expected from the AGN-unified model, that is, HEGs and BLOs are consistent -- from the point of view of their environment -- with being the same class of objects but just seen at a different inclination angle. 

\item The distributions of local galaxies' density of LEGs as well as HEGs and BLOs (now considered as a single class based on the previous result) are not statistically distinguishable. The same result applies to the comparison of FR~Is and FR~IIs.

\item The distributions of $N_{RS}$ for HEGs and BLOs  as well as LEGs, in addition to for FR Is and FR IIs, span over a wide range of richnesses.
Hence, independently of their optical and morphological classification, RGs can be found in relatively poor environments (i.e., small galaxy groups) as well as in rich galaxy clusters.
With just a few exceptions among the FR~IIs, the distributions also highlight the preference for RGs to avoid being isolated sources.

\item The radio luminosity is not connected with the source environment.

\end{itemize}

Our main result, that is to say the lack of a connection between the environment and RGs' properties (both morphological and spectroscopic), is in agreement with the findings of ~\citet{massaro19} and ~\citet{vardoulaki21}.
However, other works on RG environments had shown different results.
\citet{massaro19} discuss, in detail, possible reasons for this discrepancy  (and others). We followed a similar approach based on the results of our analysis.

The $\Sigma_5$ parameter, defined as the ratio between the number of sources and the projected area included between the central galaxy and the fifth nearest neighbor, is a widely used parameter to estimate the galaxies density and \citet{ching17} found that HEGs are associated with lower values of $\Sigma_5$ than LEGs. However, $\Sigma_5$ is a redshift-dependent quantity because faint galaxies at larger distances are undetected. The comparison of the properties of sources at a different redshift may then be inaccurate. Our method, based on redshift-independent quantities such as the galaxies' absolute magnitude, is not affected by this problem.
Furthermore, although \citet{ching17} started from a large sample, $\sim 400$ sources, their conclusion is based on a rather small group of high power HEGs.

\citet{gendre13} concluded that HEGs are found almost exclusively in low-density environments while LEGs occupy a
wider range of densities, apparently in contrast with our conclusions. However, when the comparison between LEGs and HEGs is limited to those with a FR~II morphology, they found that the richness of the two classes in indistinguishable, which is in agreement with our findings.

In conclusion, we considered sources over a wide range of redshift. Our analysis limits the effects of the different distances of the sources as much as possible; furthermore, the redshift distributions of the various subclasses are not statistically different from each other. This restricts biases introduced by the analysis of sources at different redshifts. 

However, our work considers only a relatively small group of sources and a larger sample should be used to put out results on a stronger statistical basis. Furthermore, we referred only to the local galaxies' density to characterize the RGs' environment. Other factors can be important, such as the location of a given source within its group or cluster of galaxies. This effect cannot be studied with our method due to the large contribution of interlopers in the CMDs which prevent the estimate of the location of the group or cluster center and/or the presence of close companions, for example. This requires complete spectroscopic coverage of nearby objects.

Alternatively, a more comprehensive picture of the environment could be obtained by analyzing the gaseous component in the surroundings of RGs. In particular, X-rays probe the hot gas underlying the halo hosting the RGs. Properties of this large-scale gas, such as density and temperature, in correlation with the properties of the RG (i.e., position in the cluster or group, morphology, and accretion rate) can lead to a more comprehensive description of the inter-connection between the AGN activity and the large-scale structure. The forthcoming all sky survey performed by the eROSITA instrument on board of the Spectrum-Roentgen-Gamma mission \citep{predehl21} is the ideal instrument to perform such a study. 

In the near future, the Rubin Legacy Survey of Space and Time \citep[Rubin-LSST;][]{ivezic2019} will monitor the southern sky for ten years in six filters with unprecedented depth. This will allow us to extend the study of the environmental properties of RGs to high redshifts. According to \citet{kotyla2016}, half of the RGs at $z>1$ lie in rich environments, but this is based on a small sample of 21 RGs from 3CR. 
Using their results and an elliptical galaxy template \citep{polletta2007}, we infer that Rubin-LSST will have to reach a co-added image depth of about 24 mag in the z band to make the analysis of the clustering properties up to $z \sim 0.5$ possible. This will be obtained in the first years of the survey. Instead, a mag of $\sim 26.5$ is needed to go up to $z \sim 2$, which is approximately the depth that Rubin-LSST will reach in ten years. Therefore, as the project progresses, we will be able to push the study of the RG environment from the local Universe to cosmological distances.

\section*{Acknowledgements}
The Pan-STARRS1 Surveys (PS1) and the PS1 public science archive have been made possible through contributions by the Institute for Astronomy, the University of Hawaii, the Pan-STARRS Project Office, the Max-Planck Society and its participating institutes, the Max Planck Institute for Astronomy, Heidelberg and the Max Planck Institute for Extraterrestrial Physics, Garching, The Johns Hopkins University, Durham University, the University of Edinburgh, the Queen's University Belfast, the Harvard-Smithsonian Center for Astrophysics, the Las Cumbres Observatory Global Telescope Network Incorporated, the National Central University of Taiwan, the Space Telescope Science Institute, the National Aeronautics and Space Administration under Grant No. NNX08AR22G issued through the Planetary Science Division of the NASA Science Mission Directorate, the National Science Foundation Grant No. AST–1238877, the University of Maryland, Eotvos Lorand University (ELTE), the Los Alamos National Laboratory, and the Gordon and Betty Moore Foundation.

Funding for the Sloan Digital Sky Survey (SDSS) has been provided by the Alfred P. Sloan Foundation, the Participating Institutions, the National Aeronautics and Space Administration, the National Science Foundation, the U.S. Department of Energy, the Japanese Monbukagakusho, and the Max Planck Society. The SDSS Web site is http://www.sdss.org/.

The SDSS is managed by the Astrophysical Research Consortium (ARC) for the Participating Institutions. The Participating Institutions are The University of Chicago, Fermilab, the Institute for Advanced Study, the Japan Participation Group, The Johns Hopkins University, Los Alamos National Laboratory, the Max-Planck-Institute for Astronomy (MPIA), the Max-Planck-Institute for Astrophysics (MPA), New Mexico State University, University of Pittsburgh, Princeton University, the United States Naval Observatory, and the University of Washington.


\end{document}